# Alignment free characterization of 2D gratings

*Morten Hannibal Madsen[1*], Pierre Boher[2], Poul-Erik Hansen[1], Jan Friis Jørgensen[3]*

[1]Danish Fundamental Metrology A/S, Matematiktorvet 307, 2800 Kgs. Lyngby, Denmark

[2]ELDIM, 1185 rue d'Epron, 14200 Herouville Saint Clair, France

[3]Image Metrology A/S, Lyngsø Allé 3A, 2970 Hørsholm, Denmark

*Corresponding author: mhm@dfm.dk

KEYWORDS: instrument design, scatterometry, optical characterization, inverse modelling, diffraction gratings

ABSTRACT Fast characterization of 2-dimensional gratings is demonstrated using a Fourier lens optical system and a differential optimization algorithm. It is shown that both the grating specific parameters such as the basis vectors and the angle between them and the alignment of the sample, such as the rotation of the sample around the x-, y-, and z-axis, can be deduced from a single measurement. More specifically, the lattice vectors and the angle between them have been measured, while the corrections of the alignment parameters are used to improve the quality of the measurement, and hence reduce the measurement uncertainty. Alignment free characterization is demonstrated on both a 2D hexagonal grating with a period of 700 nm and a checkerboard grating with a pitch of 3000 nm. The method can also be used for both automatic alignment and in-line characterization of gratings.

**INTRODUCTION**

Diffraction gratings are identical structures repeated multiple times within a large area leading to the optical phenomena diffraction. The properties of diffraction gratings were first described by Rittenhouse more than 200 years ago based on observations through a silk handkerchief [1]. Today, diffraction gratings are used in many applications including spectrometers [2], frequency stabilized lasers [3], and optical multiplexing devices [4]. These devices require high quality gratings and characterization is an important part of the fabrication chain. However, characterization of gratings is a time consuming task using the most common techniques such as angular diffraction, atomic force microscopy (AFM), or scanning electron microscopy (SEM). In typical setups for measuring pitch, such as diffraction in the Littrow configuration, each diffraction order is measured individually. This gives an uncertainty as low as 15 pm for a 144 nm grating [5]. However, alignment of the sample is very critical for these types of measurements and it only measures one axis at a time. In other words, it requires two measurements to obtain both axis of a 2D grating. With AFM and SEM one obtains an image of the grating and an average period can be found from subsequent image processing. It is challenging to perform traceable measurements with an SEM, whereas traceable AFM measurements can be performed using interferometers on the axes or transfer standards. Expanded uncertainties are around 0.3 nm and 1 nm for AFM and SEM, respectively [6].

In this paper a fast method for characterization of periodic gratings is presented. Focus is on 2D gratings, an example is sketched in Fig. 1, but the proposed method works equally well on 1D gratings. However, the data acquisition time is the same for both 1D and 2D grating, and hence the improvement in measuring time is greatest for the 2D grating. Bi-periodic gratings have previously been imaged using a Fourier lens system [7], but no measurements of the grating pitches were made.

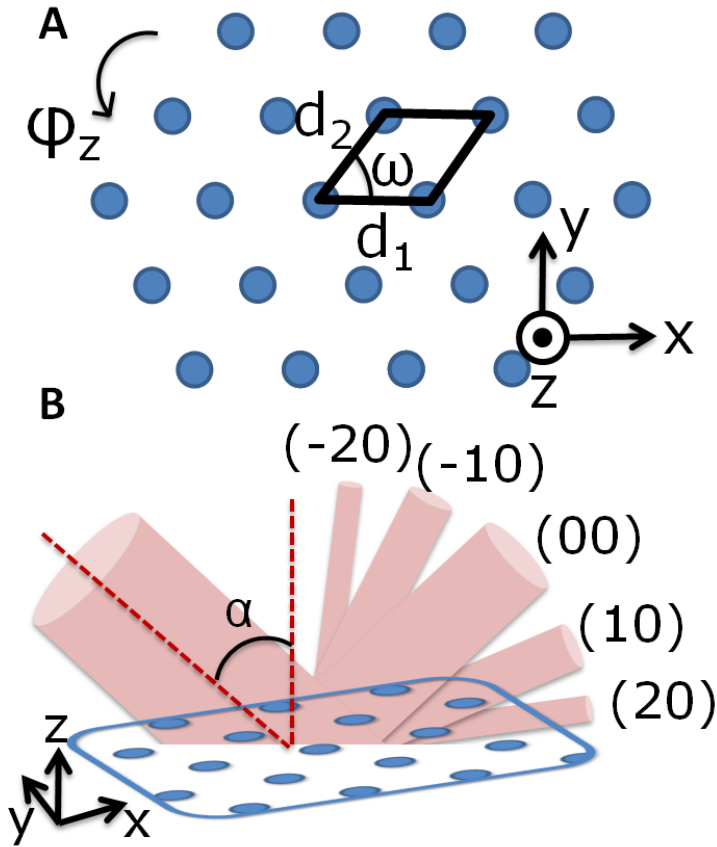

Figure 1. Sketch of a 2D grating. (A) The grating parameters for pitches $d_1$ and $d_2$ and the angle of the unit cell, $\omega$, are indicated on the figure. The rotation of the sample around the z-axis, $\varphi_z$, is indicated. The alignment parameters $\varphi_x$ and $\varphi_y$ is the rotation in counter-clockwise direction around the x- and y-axis, respectively. (B) Sketch of diffraction on a 2D hexagonal grating. Some examples of labelling using Miller indices are indicated in the figure. The angle of the incoming light, $\alpha$, is indicated on the figure.

Topological parameters for a diffraction grating can be found using the scatterometry technique [8][add ref.]. In scatterometry the diffraction efficiencies, typically as a function of either diffraction angle or wavelength, are used as a unique 'fingerprint' for a given surface. Parameters that describe the surface are then found using an inverse modelling approach, where simulated diffraction efficiencies are matched to the experimental found values. Angular scatterometers provide high resolution measurements of the diffraction efficiencies [9–11], but measurements are time-consuming and alignment of the sample is cumbersome. Furthermore they can only measure diffraction in one direction, thus characterization of 2D gratings require two measurements. The hemisphere detector setup presented in this paper can also be used for scatterometry measurements, but this requires information of the orientation of the sample. Thus, the alignment-free method presented here is an important step for automatic scatterometry measurements using a hemisphere detector. So far scatterometry has mostly been used in the semi-conductor industry, but with a fully automatic scatterometer that can characterize both 1D and 2D gratings; the scatterometry technique will be interesting for many new industry sectors. This can include the injection molding industry with smart functionalization of plastic [12] and the roll-2-roll industry [13].

**THEORY**

Calculation of the position of diffraction spots is used in e.g. X-ray scattering experiments [14] and electron beam microscopes [15]. In this paper the positions of diffraction spots for 2D gratings are calculated for a screen in the back focal plane. We follow the approach described in Ref. [16] where the specific case of imaging the diffraction spots on a semi-transparent screen is derived. Here the Laue condition $\Delta \vec{k} = \vec{G}$, where $\Delta \vec{k}$ is defined as the difference between the incoming wave vector $\vec{k}$, and the diffracted wave vector $\vec{k}'$, and $\vec{G}$ is the reciprocal lattice vector, is fulfilled for the diffraction spots only.

We define the basis vectors of the grating to

$$\vec{a}_1 = d_1 \begin{pmatrix} 1 \\ 0 \\ 0 \end{pmatrix} \text{ and } \vec{a}_2 = d_2 \begin{pmatrix} \cos\omega \\ \sin\omega \\ 0 \end{pmatrix}$$

with grating parameters defined in Fig. 1A. The simplest vector perpendicular to both $\vec{a}_1$ and $\vec{a}_2$ is $\vec{a}_3 = (0,0,1)$. Rotation and tilt correction of the sample is handled using rotational matrices around the x-, y-, and z-axis. In the first steps of the optimization algorithm no rotational misalignments are assumed, but in the last steps a full correction of the rotational misalignments is applied.

The reciprocal lattice vectors are then defined by

$$\vec{b}_1 = 2\pi \frac{\vec{a}_2 \times \vec{a}_3}{\vec{a}_1 \cdot \vec{a}_2 \times \vec{a}_3},$$

$$\vec{b}_2 = 2\pi \frac{\vec{a}_3 \times \vec{a}_1}{\vec{a}_1 \cdot \vec{a}_2 \times \vec{a}_3}.$$

Relative to the sample the incoming light is given by

$$\vec{k} = \frac{2\pi}{\lambda} \begin{pmatrix} \sin(\alpha - \varphi_y) \\ \sin(-\varphi_x) \\ -\cos\alpha \end{pmatrix},$$

where α is the angle of the incoming beam and λ is the wavelength of the incoming beam. The scattering vector is now calculated using

$$\vec{k}' = \frac{2\pi}{\lambda}\left(h\vec{b}_1 + k\vec{b}_2\right) + \vec{k}.$$

The coordinates of the scattering vector can directly be converted to polar coordinates, which is the measured quantity using the Fourier lens system.

**EXPERIMENTAL SETUP**

For the measurements an experimental setup based on a Fourier lens system (EZContrast from ELDIM), also referred to as a conoscope system was used. This type of system has first been introduced commercially by ELDIM in 1993 [17]. Please note the difference to Fourier scatterometers that collects light through a conventional high numerical aperture objective and are typically based on a interferometric setup [18–20].

The used Fourier lens system collects quasi all light coming from one spot on the sample surface (maximum incidence angle up to $88^{o)}$ and project it to a first Fourier plane. One point on this Fourier plane corresponds to one light direction. This first Fourier plane is reimaged on an imaging sensor via a field lens and an imaging lens. In between an iris in direct view of the sample surface defines the measurement spot independently of the angular aperture. Three motorized filter wheels with 31 narrow band-pass filters (typical band width of less than 8 nm) were placed in the beam path in front of the CCD sensor as sketched in Fig. 2A. In this part of the optical setup the system is quasi telecentric to ensure fixed transmittance for all the imaging fields. Illumination for reflective samples is made using a beam splitter and an additional illumination lens to build an illumination Fourier plane. The beam splitter is positioned between the field lens and the iris, and can be removed if transparent back-lid samples are investigated. Homogenous illumination was achieved using a stabilized halogen lamp and an integrating sphere and masks were made for illumination of multiple individual spots simultaneously. We used masks with up to 15 holes thus giving 15 illumination rays with same azimuth angle but a polar angle ranging from 0° to 70° in steps of 5°.

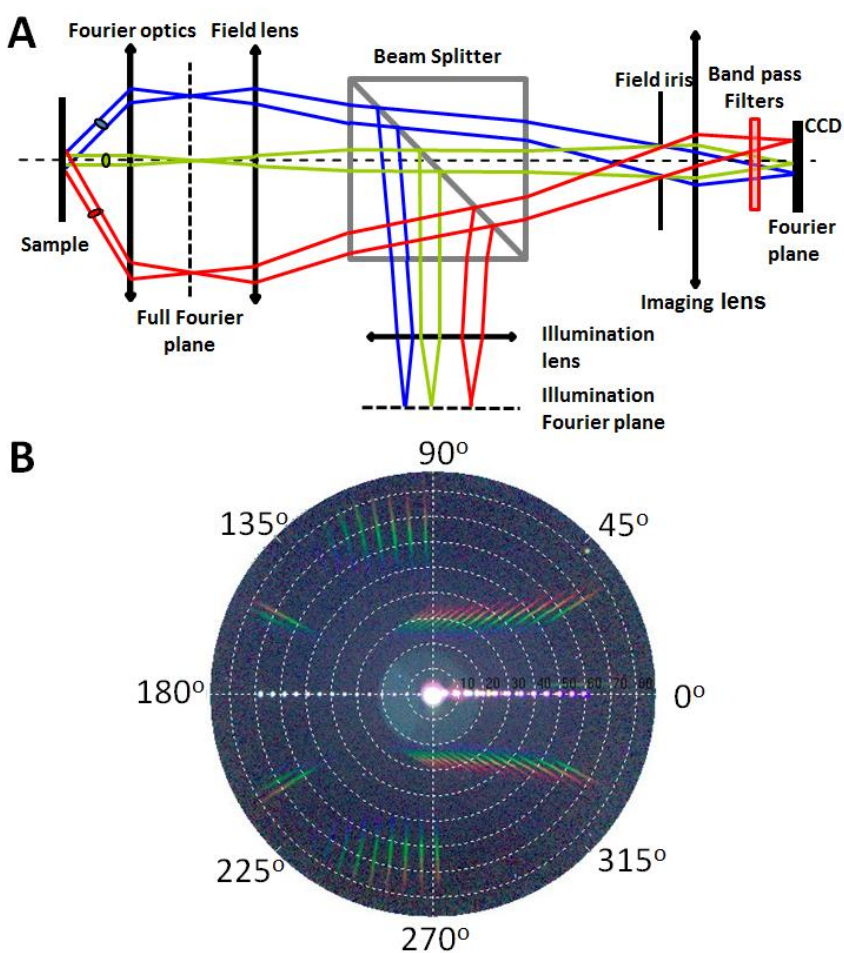

Figure 2. Fourier lens imaging. (A) Sketch of the basic principle of a Fourier lens system. Light is coupled in using the beam splitter positioned between the field lens and the iris. (B) Experimental data of a multi-wavelength image on a 700 nm hexagonal grating.

The Fourier lens system was mounted on a motorized stage perpendicular to the sample. The sample was then placed under the Fourier lens system and brought into focus by adjusting the height of the system (*z*-direction). The measurement spot was observed directly using additional optics and a camera inside the Fourier lens system. A more detailed description of the experimental setup can be found in Refs. [7,21].

For each band pass filter an image was obtained with an average acquisition time of 2 sec. All the images were stored in a multi-spectral file and these can be plotted to a single image as shown in Fig. 2B. The positions of the diffraction spots were extracted and from theses the average pitches can be calculated. However, this requires information of the alignment of the sample. For robust and alignment-free calculation of the pitches a software algorithm was used for optimization for both the grating and alignment parameters.

**OPTIMIZATION ALGORITHM**

A differential evolution algorithm [22] has been used to find the optimal set of parameters from the model to fit the experimental data. Parameters relating to both the grating and to the alignment of the sample are optimized simultaneously. Specifically we optimize for the grating pitches, $d_1$ and $d_2$, the angle between the lattice vectors, $\omega$, and for rotation of the sample around the x-, y- and z-axis, $\varphi_x$, $\varphi_y$ and $\varphi_z$. The used differential evolution algorithm can be found at [23] and is based on the implementation described in Ref. [24]. The value-to-reach (VTR) specifies a condition in the program for when to stop the optimization. The VTR was set to 5 and $1·10^{-6}$ when finding the Miller indices and all the grating parameters, respectively. The huge difference illustrates the point that only a rough estimate is necessary when the Miller indices are determined. For optimization of the parameters the calculations were most often stopped when the maximum number of iterations was reached, as will be discussed below.

The optimization algorithm data have been validated by simulating data for a 3000 nm square lattice with different level of noises added to both the azimuth and polar angles. The noise is generated using randomly drawn numbers from a normal distribution with standard deviation $\sigma_{noise}$. For generating the data, Miller indices ranging from -10 to 10 and multiple incoming illumination rays along a single azimuth from 5º to 65º in steps of 5º, giving a total of 5733 diffraction spots, were used. In the absence of noise the reconstruction of the simulated data using the differential evolution algorithm gives a relative accuracy of the six parameters of less than $10^{-5}$. In Fig. 3 the deviation from the reconstructed value to the input parameters is plotted with respect to the amount of added noise. For a noise level of up to $\sigma_{noise} = 5º$ on both azimuthal and polar angle of the diffraction spots the reconstruction still gives results with a deviation of less than 0.2 % for the grating lattice parameters. For all calculations 18000 iterations were used. This number was found from evaluation of the dataset with the highest added noise level as shown in the insert in Fig. 3.

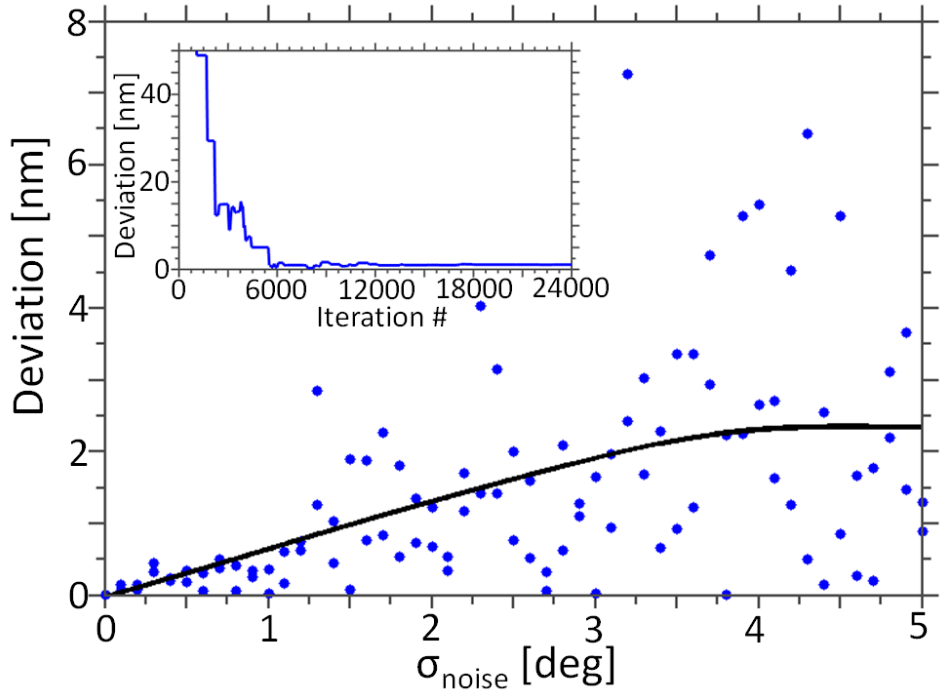

Figure 3. Reconstruction of simulated data for a 3000 nm 2D grating. The deviation from the target value is plotted as a function of added noise for both axes at the same time. The noise is added to both the polar and azimuthal angles of the data. The solid line is a guide to the eye. The insert shows the deviation as a function of the number of iterations made with an added noise of $\sigma_{noise} = 5°$.

For experimental data a number of challenges arise. First, the angle of the incoming light should be known. Second, overlapping diffraction spots should be treated carefully in the data analysis and i.e. be removed. And third, all diffraction spots have to be labeled with their respective Miller indices. The workflow for processing experimental data is sketched in Fig. 4.

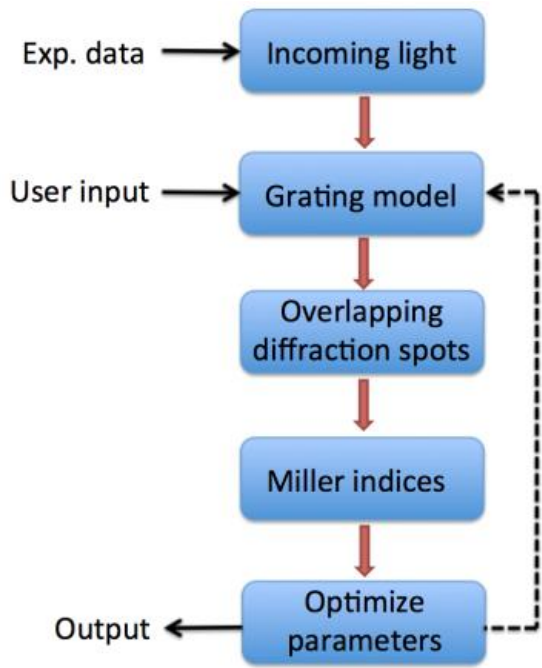

Figure 4. Workflow for using the optimization algorithm on experimental data. After the 'Miller indices' step the user has a list of the position of each diffraction spot with their corresponding Miller indices and incoming wave vector. The final output is values for the grating and alignment parameters, with typically the grating pitches as the most important ones.

For the first challenge, the angle of incidence is defined by the holes in the illumination mask, and they can be measured directly from the specular reflections. It is assumed that the illumination stay constant during a measurement. The second challenge, the overlapping diffraction spots, can for instance be removed from calculations using *a priori* knowledge of the approximately grating pitch. The third challenge, the labeling, is more challenging and we also used a differential evolution optimization approach with discrete values of the Miller indices for each measured diffraction spot. This of course also requires some *a priori* knowledge of the system and grating parameters. It might seem to as a harsh condition to have *a priori* information of the sample to investigate, but typically one knows the design specification of the sample, which is sufficient at this stage of the optimization. Alternatively a high resolution optical image of the grating can be used as the initial input for the grating model. In some cases it might be necessary with an iterative process with several optimization steps of the Miller indices and grating parameters, as indicated with a dashed arrow in Fig. 4.

**EXPERIMENTAL DATA AND DISCUSSION**

Two types of 2D gratings have been measured with the Fourier lens system; a hexagonal grating with a nominal pitch on both axes of 700 nm, and a checkerboard grating with nominal pitches of 3000 nm. For reference measurements of the 700 nm grating a metrology AFM (NX20, Park Systems) was used. The traceability of the AFM was achieved using transfer standards according to the procedure described in Ref. [25]. The 3000 nm grating was a transfer standard with certified reference values measured by an angular diffraction technique. The reference values are listed in Tab. 1.

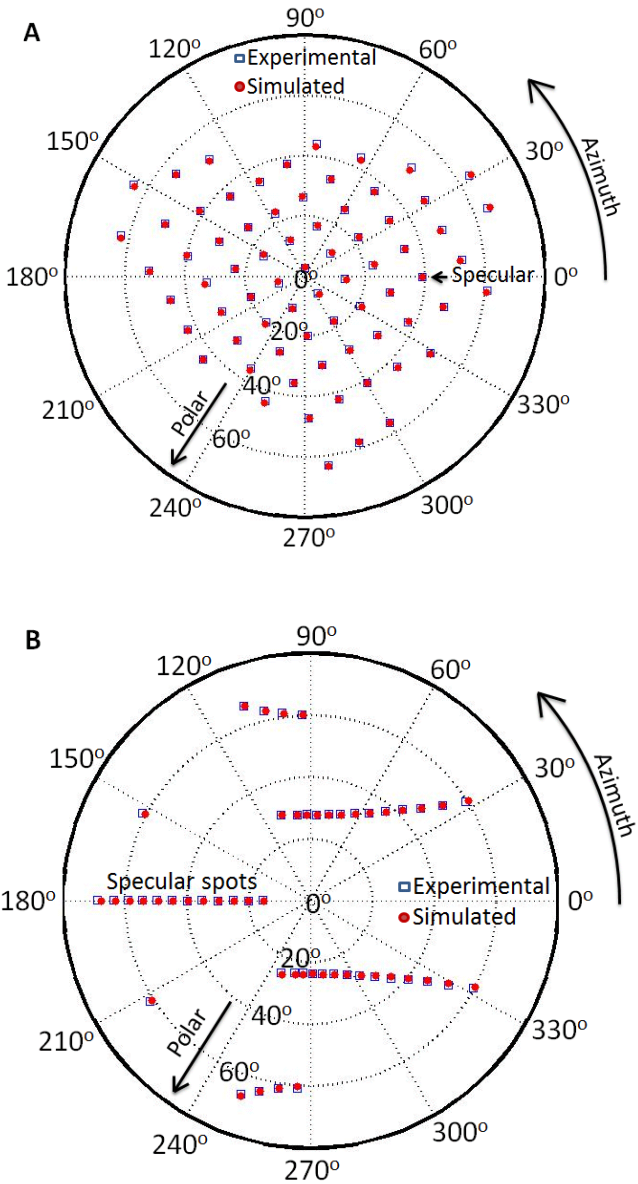

Figure 5. Experimental data (blue squares) and best fit (red circles) for measurements using the Fourier lens system and the optimization algorithm. (A) 2D checkerboard grating with an angle of incidence of 40° using a wavelength of 527 nm. (B) 2D hexagonal grating with a period of 700 nm for multiple incoming light rays.

The 3000 nm grating was illuminated with a single incident beam at a polar angle of 40° in order to avoid overlapping diffraction orders, as shown in Fig. 5A. For the 700 nm hexagonal grating a total of 14 points were illuminated simultaneously, as shown in Fig. 5B. The points were on the same azimuth with approximately 5° polar angle spacing.

The peak positions were measured at the center of the diffraction spots and the optimization algorithm described previously used to calculate the grating and alignment parameters. The found grating parameters and their estimated expanded uncertainty are listed in Tab. 1.

**Table 1. Reference measurements of the gratings are performed using traceable atomic force microscopy (AFM) and angular diffraction for the hexagonal and square lattice, respectively. $U_{Ref}$ and $U_{FL}$ denotes the expanded standard uncertainty (k=2) for the reference and Fourier lens system measurements, respectively.**

|  | Parameter | Reference | $U_{Ref}$ | Fourier lens | $U_{FL}$ |
|---|---|---|---|---|---|
| hexagonal | $d_1$ | 699.5 nm | 2.3 nm | 700 nm | 12 nm |
|  | $d_2$ | 700.5 nm | 2.3 nm | 703 nm | 12 nm |
|  | $\omega$ | 60.3° | 2.0° | 59.9° | 0.5° |
| square | $d_1$ | 3001.1 nm | 3.0 nm | 3001 nm | 15 nm |
|  | $d_2$ | 3001.9 nm | 3.0 nm | 3004 nm | 15 nm |
|  | $\omega$ | 89.97° | 0.08° | 90.1° | 0.5° |

The uncertainties in Tab. 1 are given at the 95 % confidence limit (k=2). The uncertainties for the Fourier lens system are estimated using.

$$U_{FL} = 2\sqrt{u_{sys}^2 + u_{sam}^2 + u_{rep}^2 + u_{spot}^2}$$

where the four elements, $u_{sys}$, $u_{sam}$, $u_{rep}$, and $u_{spot}$, represents the uncertainties associated with the system, the sample, the reproducibility, and the spot detection, respectively. The angle accuracy of the Fourier lens system is less than 0.2º [21], leading to an estimated uncertainty of the system to $u_{sys}$ = 2 nm and $u_{sys}$ = 3 nm for the 700 nm and 3000 nm grating, respectively. The uncertainty contribution from uniformity and alignment of the sample, $u_{sam}$, is negligible, as the samples are of very high quality and the alignment is corrected for using the algorithm described in this paper. For evaluating the reproducibility 9 measurements of the same sample was acquired, with the sample removed and repositioned in between the measurements. The reconstructed grating and system parameters for the reproducibility study of the 2D checkerboard grating with a period of 3000 nm are plotted in Fig. 6. The standard deviation for the pitches is calculated to $\sigma$ = 5 nm, and this is used as the uncertainty contribution from reproducibility, $u_{rep}$ = 5 nm. The reconstruction of the system dependent parameters, see Fig. 6B, show that $\varphi_x$ and $\varphi_y$ are almost constant, as the sample is leveled. The sample is rotated differently for each measurement and hence $\varphi_z$ varies with several degrees. It should be noted that with the used reconstruction algorithm, the grating parameters become independent of the system parameters. In other words, the rotation of the sample is negligible for the reconstruction of the grating pitches and angle.

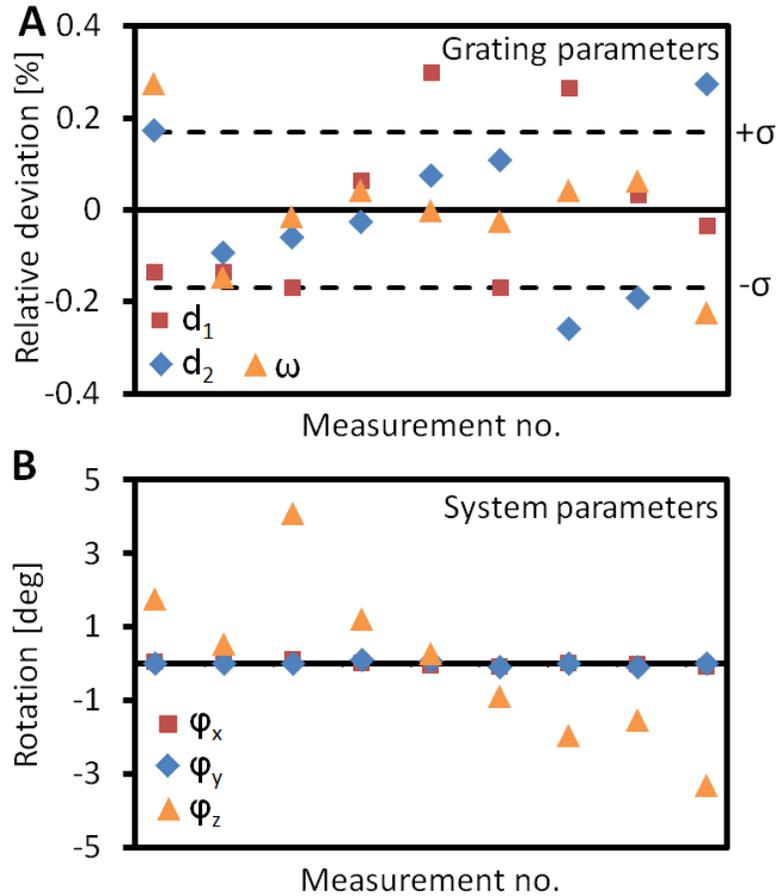

Figure 6. Reproducibility study of the system using the 3000 nm 2D checkerboard grating. A total of 9 measurements were obtained and reconstructed, where the sample was removed and repositioned between all measurements. (A) The relative deviation of the reconstructed grating parameters. The deviation is the difference between the average value and the found value for the specific measurement. The standard deviation of the reconstructed pitches, ±$\sigma$, is indicated with dashed lines and is equal to $\sigma$ = 5 nm. For each measurement has two values been extracted; one for each axis of the grating. (B) Reconstruction of the system parameters, that is, rotation around the x-, y- and z-axis. The sample is leveled and hence $\varphi_x$ and $\varphi_y$ are nearly constant for all the measurements. On the other hand $\varphi_z$ varies many degrees between each measurement.

The uncertainty of the spot detection is estimated to give uncertainties $u_{spot}$ = 3 nm and $u_{spot}$ = 5 nm for the short and long period gratings, respectively, based on the difference in applying different methods for the spot recognition. As non-coherent light is used, typically spanning a wavelength range of 8 nm, the spot sizes increase. By switching to a frequency stabilized laser source instead, this uncertainty contribution can almost be neglected. It will nevertheless be necessary to break the coherence of the light to avoid coherence effects in the imaging system.

A critical step in the workflow is the unique detection and labelling of all diffraction spots, especially for the setup with multiple simultaneous illumination angles. More accurate *a priori* knowledge of grating parameters is needed when diffraction spots are closer to each other and a lower value for when to stop the optimization is needed, thus increasing the simulation time. As an example, the diffraction spots in Fig. 5B are much closer than the more uniformly distributed diffraction in Fig. 5A.

**OUTLOOK AND SUMMARY**

The presented system is much faster than conventional angular scatterometers, especially for 2D gratings, as all diffraction spots are measured simultaneously. The precision and accuracy of the measurements is better with a conventional angular scatterometer, but for applications where an expanded uncertainty of around 10 nm is acceptable, the described setup and automatic alignment algorithm is superior.

The next step will be to use the measurements for scatterometry analysis, where one can reconstruct the topographic parameters of the surface structures, e.g. height and width. In the modeling of scatterometry data the period is often kept fixed to reduce the number of fitting parameters [26,27]. By measuring the periods directly and simultaneously align the sample; we expect to improve the reconstruction part of scatterometry.

In summary we have shown that we can characterize 2D diffraction gratings with a single measurement using a Fourier lens system. By using a differential evolution algorithm, parameters for both the grating and the alignment of the sample are optimized simultaneously. Thus, the basic parameters of gratings can be measured without careful alignment of the sample. More specifically, the lattice vectors and the angle between them have been measured, while the corrections of the alignment parameters are used to improve the quality of the measurement, and hence reduce the measurement uncertainty.

**Funding.**  Eurostars (E8875 InFoScat); EMPIR (14IND09 MetHPM)

**Acknowledgment**. We thank NIL Technology ApS (Diplomvej 381, 2800 Kgs. Lyngby, Denmark) for providing the high quality hexagonal grating. This project has received funding from the EMPIR programme co-financed by the Participating States and from the European Union's Horizon 2020 research and innovation programme through the project 14IND09 MetHPM. We also acknowledge the financial support from Eurostars project E8875 - InFoScat.